\documentclass[twocolumn, times]{aastex701}

\usepackage{graphicx}
\usepackage{epstopdf}
\usepackage{amsmath}
\usepackage{placeins}
\usepackage{multimedia}
\usepackage{natbib}
\usepackage{enumitem}

\shortauthors{Liu et al.}
\shorttitle{}

\begin{document}

\title{Bidirectional plasma jets driven by magnetic reconnection: observations by GST and SDO}

\author{Yangyu Liu}
\affil{Xinjiang Astronomical Observatory, CAS, Urumqi, 830011, People's Republic of China, shenjh@xao.ac.cn, yangxu@xao.ac.cn}
\affil{School of Astronomy and Space Science, UCAS, Beijing 100049, People’s Republic of China}
\email{liuyangyu@xao.ac.cn}
\author{Jinhua Shen}
\affil{Xinjiang Astronomical Observatory, CAS, Urumqi, 830011, People's Republic of China, shenjh@xao.ac.cn, yangxu@xao.ac.cn}
\email{shenjh@xao.ac.cn}
\author{Xu Yang}
\affil{Xinjiang Astronomical Observatory, CAS, Urumqi, 830011, People's Republic of China, shenjh@xao.ac.cn, yangxu@xao.ac.cn}
\email{yangxu@xao.ac.cn}
\author{Shuai Gu}
\affil{Xinjiang Astronomical Observatory, CAS, Urumqi, 830011, People's Republic of China, shenjh@xao.ac.cn, yangxu@xao.ac.cn}
\affil{School of Astronomy and Space Science, UCAS, Beijing 100049, People’s Republic of China}
\email{gushuai@xao.ac.cn}
\author{Jianping Li}
\affil{Key Laboratory of Dark Matter and Space Astronomy, CAS, Nanjing 210023, People's Republic of China}
\email{jpli@pmo.ac.cn}
\author{Haisheng Ji}
\affil{Key Laboratory of Dark Matter and Space Astronomy, CAS, Nanjing 210023, People's Republic of China}
\email{jihs@pmo.ac.cn}

\begin{abstract}
Using high-resolution photospheric and chromospheric observations taken by the Goode Solar Telescope (GST), we studied  two groups of bidirectional plasma jets occurring in active region NOAA 13110. Supplementary observations are also provided by Solar Dynamics Observatory’s (SDO) Atmospheric Imaging Assembly (AIA) and Helioseismic and Magnetic Imager (HMI). From the photospheric observations and magnetograms, the two successive bidirectional plasma jets were initially located in the vicinity of the polarity inversion lines and at the intersection of the umbra and penumbra of the sunspot, followed by magnetic flux emergence and cancellation. As the cool filamentary threads are continuously emerging from the lower chromosphere and interact with overlying horizontal magnetic loops, it leads to the bidirectional jets, erupting outflow plasmoids, and heating coronal magnetic loops. We find that the bidirectional jets extended from the central excitation location in opposite directions, at the speed of about dozens of ~\text{km~s$^{-1}$}. For the first jet, the initial brightening first appears in 304~\AA{}, about 30 s earlier than the H$\alpha$ observations, indicating that magnetic reconnection takes place in the transition region. While the initial reconnection for the second jet occurs simultaneously in H$\alpha$ and 304~\AA{}, showing the recurrent eruptions. These observations confirm that the bidirectional plasma jets can be generated by magnetic reconnection between the rising filamentary threads or material and the overlying horizontal magnetic loops. Our results provide new insights into the generation of the bidirectional plasma jets and reconnection-based coronal heating.
\end{abstract}

\section{Introduction} 

Magnetic reconnection is considered to be the process of sudden release of free magnetic energy in the solar atmosphere. It plays a key role in powering different scales of solar activities, such as solar flares, coronal mass ejections (CMEs) and jets/surges \citep{priest2000method}. With the application of high spatiotemporal resolution solar instruments, more and more observations have reported that the jets or surges are driven by magnetic reconnection from the lower atmosphere to the corona, producing H$\alpha$ surges \citep{roy1973,asai2001}, as well as extreme ultraviolet (EUV) and X-ray jets \citep{chae1999, cheung2015,sterling2015,shen2018}. According to the emerging-reconnection model, \cite{shibata1994a} divided coronal jets into the anemone jets and two-sided loop jets, the anemone jets could be produced by the interaction between bipolar emerging flux from below the solar surface and vertical or oblique coronal fields, while the two-sided loop jets are believed to be due to the magnetic reconnection between the bipolar emerging flux and the overlying horizontal coronal fields, and they proposed the types of jets generated entirely depend on the morphology structure of the surrounding magnetic fields. Based on the magnetic morphology of coronal jets, \cite{moore2010} classified the coronal jets into standard jets (inverted Y-shaped) and blowout jets. For the blowout jets, the minifilament with shear triggers the coronal reconnection at a nullpoint, showing an brightening inside the base arch and an extra jet spire near the outside bright point. For large-scale coronal jets, \cite{moore2018} further reported that internal tether-cutting reconnection under the erupting miniﬁlament ﬂux rope drives the miniﬁlament’s rise and external breakout reconnection produces erupting jets, associated with energetic flares and CMEs. Therefore, the study of the production of plasma jets provides a useful diagnostic of spatial location where magnetic reconnection occurs.

Using the 2D magnetohydrodynamic (MHD) simulations, \cite{yokoyama1995} and \cite{yokoyama1996} successfully modeled the anemone jets and the two-sided loop jets. For the two typical jets, the anemone jets show an inverted Y-shaped or $\lambda$ structure, and followed by a straight plasma beam as well as a jet-base bright point like-dome. The anemone jets are widely studied through observations and simulations \citep{chae1999, zhang2014, moore2018, ni2017, wyper2018}. They often take place in the light bridge of sunspots \citep{yang2015, tian2018}, penumbra \citep{zeng2016, reid2018}, and the mixed-polarity region near the sunspots \citep{jiang2007,liu2016}. On the contrary, the bidirectional plasma jets move away from the reconnection site in opposite directions, which were first reported in EUV spectra and coronal X-ray images respectively \citep{innes1997, alexander1999}, showing Doppler shifts changing from red to blue in a few seconds. An anemone or bidirectional jet is believed to be an indirect evidence of magnetic reconnection from the chromospheric jets to coronal jets \citep{mulay2016, liu2016, tian2018, kumar2023,joshi2024}.

According to the report by \cite{sterling2015}, the successful and failed eruptions of a minifilament could lead to blowout jets and standard jets respectively. On the other hand, with high-resolution observations from ground and space, some observations reported that minifilament eruption plays an important role in driving the anemone jets and two-sided loop jets \citep{sterling2019, shen2018}. \cite{raouafi2010micro} reported that the twisted filaments will be powered coronal jets and recognized as a progenitor of coronal jets. According to the simulations of a coronal jet driven by filament ejections \citep{wyper2017, wyper2018}, a large-scale jet is caused by magnetic reconnection between the rising magnetic ﬂux rope and the external ﬁeld as the rising minifilament erupts, leading to the breakout reconnection that removes the restraining field and generates outflow of plasma and internal reconnection that forms a flare current sheet. Some observations and simulations found that magnetic ﬂux cancellation at polarity inversion lines (PILs) under the miniﬁlament builds up the free energy and triggers coronal jets in quiet regions, coronal holes, and active regions \citep{panesar2016,sterling2016, panesar2018}. Therefore, the relationship between eruptive minifilaments and coronal jets is still an open question.

Bidirectional jets (also named two-sided loop jets) exhibit two sets of erupting plasma beams from the reconnection region in opposite directions, most bidirectional jets are closely related to filament eruptions \citep{sterling2019,yang2019}. \cite{sterling2019} reported that the erupting minifilament inside of flux rope interacts with overlying horizontal magnetic field at an altitude of approximately 30000 km, leading to the heating at reconnection location and generating the two-sided eruptive plasma jets, they thought that two-sided loop jets are driven by the eruptive minifilament due to magnetic cancellation at PILs \citep{shen_Stereoscopic_2019, yan2021, yang2024}. \cite{tan2023} presented that bidirectional jets are formed by the magnetic reconnection between the ﬁlament and emerging loops in ﬁlament channel. \cite{tian2017} reported that bidirectional jets are due to interact with two small-scale adjacent ﬁlamentary threads, and accompanied by magnetic flux emergence and cancellation in a mixed-polarity region near the sunspots. \cite{yang2019} reported that the two recurrent bidirectional plasma jets are the result of the interaction between minifilament and large-scale filament. The continuous magnetic cancellation at PILs lead to the accumulation and rapid release of energy throughout the filament channel. In addition, \cite{jiang2013} reported that the evolution of magnetic fields in region with bidirectional jets shows continuous emergence, they thought that recurrent bidirectional jets are caused by the magnetic reconnection between the continuous emerging bipolar and the overlying horizontal magnetic fields. Some observations also support this kind of emerging flux model \citep{kundu1998SoPh, zheng2018}. However, compared with a large number of reported anemone jets, the observed study of bidirectional jets is still rare.

Recently, using the high-resolution and multiwavelength observations, \cite{antolin2021} reported that nanojets are regarded as the signature of reconnection-based coronal heating from misaligned magnetic field lines at small angles. \cite{tang2025} reported that the nanoflare generated by post reconnection loops forms downstream with bidirectional jets, indicating the small-scale and large-scale reconnection systems are similar. \cite{chen2025} represented that the continuous outflowing bright blobs are caused by the reconnection magnetic loops with a twisted and braided at small angle, similar to the nanojets reported by \cite{antolin2021}. However, up to now, a large amount of small-scale eruptions only show the localized brightening (such as nanoflares), but the energy transmission process cannot be well confirmeddue to the limitations of the observations. Therefore, the high-resolution and multiwavelength observations are very important for understanding the driving mechanism of bidirectional jets and reconnection-based coronal heating.

  
Using high-spatiotemporal-resolution observations of GST, we reported the two groups bidirectional jets occurring in the active region NOAA 13110 on October 2, 2022. The bidirectional plasma jets are powered by continuously emerging filamentary threads or filament material interacting with overlying horizontal magnetic loops. And the brightest position as a central excitation is observed by the H$\alpha$ line core (6562.8~\AA{}) and wing, and EUV. For this active region, \cite{yang2025} and \cite{song2023} reported separation motion of the white light flare ribbons and spectral features of the white light kernels, respectively. In this paper, we focus on the bidirectional erupting plasma jets and provide a clear insight into the magnetic reconnection process using high-resolution observations. Section 2 describes instruments and data. In Section 3 we present our observational results. We summarize and discuss our findings in Section 4.

\section{Observations} 

On October 2nd, 2022, two groups of bidirectional jets were observed before the onset of an X1.0 class flare (SOL20221002T19:58) at the edge of the leading sunspot in NOAA active region 13110. 
 
High-spatial-resolution observations are performed with the Goode Solar Telescope (GST, \cite{goode2012SPIE}). The Visible Imaging Spectrometer (VIS/GST, \cite{cao2010SPIE}) is based on a single Fabry-Perot etalon that produces a narrow 0.08 ~\AA{} bandpass. VIS can provide solar chromosphere observation at a scale of $0.029^{\prime\prime}$/pixel over a $74^{\prime\prime} \times 68^{\prime\prime}$circular field of view (FOV) at the H$\alpha$ 6563~\AA\ line. In this study, VIS performs imaging spectroscopy at seven wavelengths, including the H$\alpha$ line-core (6562.8~\AA{}) and the off-bands ($\pm$0.4~\AA{}, $\pm$0.6~\AA{}, $\pm$0.8~\AA{}). The Broadband Filter Imager (BFI/GST, \cite{cao2010SPIE}) records the solar photosphere images with a 10~\AA{} bandpass filter centered at TiO 7057~\AA{} in a 60\arcsec $\times$ 60\arcsec~field of view. The resulting images have a scale of $0.03^{\prime\prime}$/pixel and a cadence of 15~s. The GST observations in TiO band and H$\alpha$ provide subarcsecond spatial resolution.

GST is powered by a adaptive optics system that utilizes a wavefront sensor with 308 sub-apertures across the telescope pupil. BFI/TiO images and VIS/H$\alpha$ images are speckle reconstructed to reduce telluric seeing distortions using the KISIP7 code (Kiepenheuer Institute Speckle Interferometry Package, \cite{woger2007field}). 

The Atmospheric Imaging Assembly (AIA; \cite{lemen2012}) images in wavelengths of 94, 131, 171, 193, 211, 304, 335, and 1600~\AA{} are used to trace the evolution of the bidirectional jets. The temperature response of AIA images ranges from 0.06 to 20 MK, covering from the lower chromosphere to the corona. Differential emission measure (DEM) analysis \citep{cheung2015,su2018determination} utilizes six AIA EUV channels (94, 131, 171, 193, 211, and 335~\AA{}) to determine plasma temperatures and densities, the temperature range is from  0.3 to 10 MK for our analysis. Line-of-sight (LOS) magnetograms and vector magnetic field from the Helioseismic and Magnetic Imager (HMI; \cite{scherrer2012}) are used to track the evolution of the LOS magnetic field and the horizontal field in the region where bidirectional jets occur.

The data set from 18:15 UT to 19:20 UT was selected for this study. All imaging data are registered and aligned with the reference HMI continuum intensity image at 18:30 UT, following the algorithm described by \cite{yang2022}.

\begin{figure*}[htbp]
  \centering
  \includegraphics[scale=0.4]{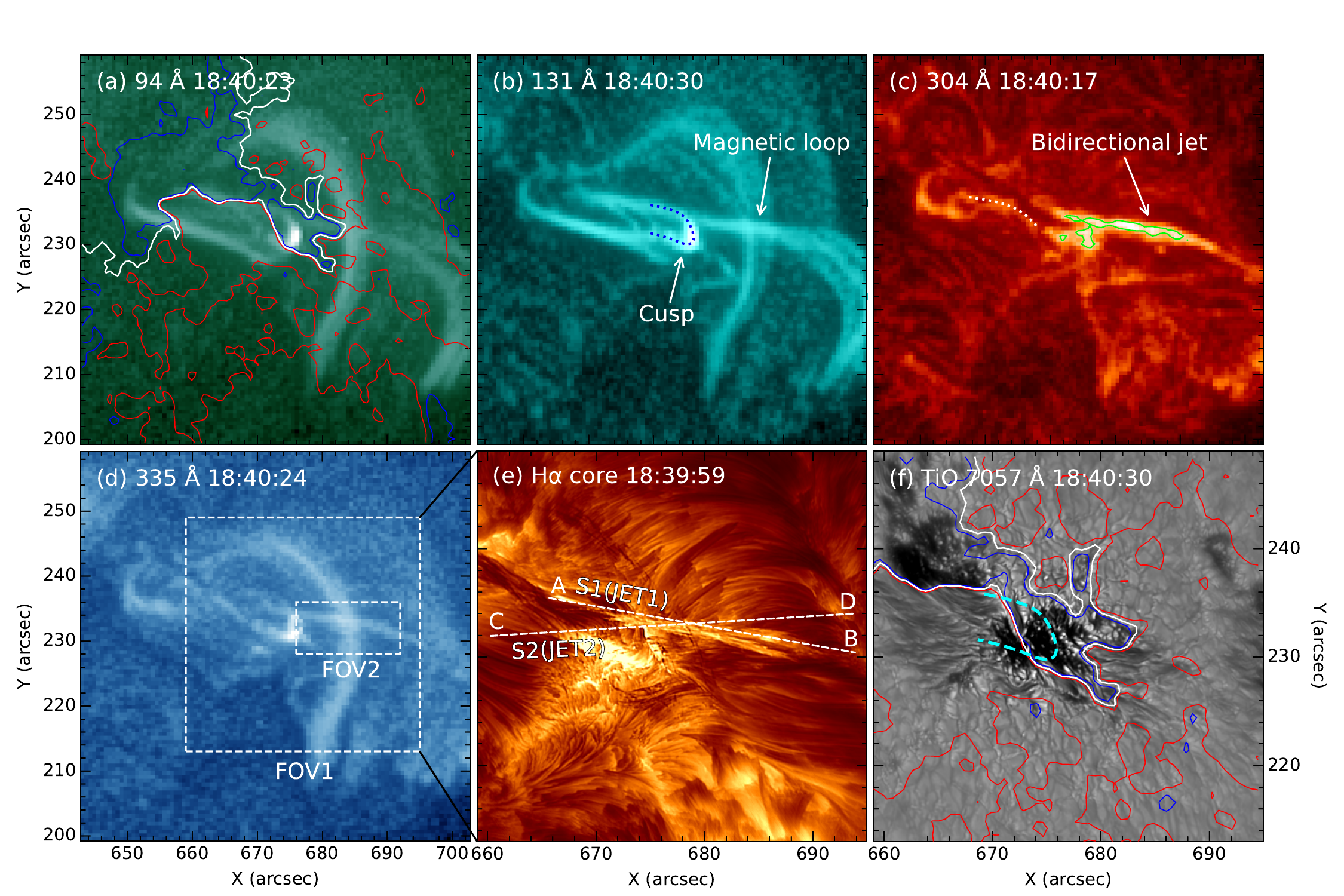}
  \caption{Panels (a--d): A bidirectional jet in 94, 131, 304, and 335~\AA{} images from AIA observations. Panels (e)--(f): The chromospheric and photospheric images taken by GST in H$\alpha$ line core and TiO band, respectively. The contours with the value of 100, 150 G (red: positive field, blue: negative field) are overlaid on background images in panels (a) and (f), the white lines indicate the PILs. The white dashed lines S1 (along AB for JET1) and S2 (along CD for JET2) in panel (e) give the positions marked as space-time diagrams in Figure 5 and 6. The two boxes (FOV1 and FOV2) in panel (d) respectively correspond to panels (e-f) and the region where the jet's flux was calculated for Figure 5 (a1-b1).  An animation of panels is available online, showing the two groups of bidirectional jets marked as JET1 and JET2. It covers the evolution from 18:15 UT to 19:20 UT. The video duration is 10 seconds. (An animation of this figure is available online.)
  }
  \label{f1}
  \movie[width=0.8\textwidth,showcontrols]{}{f1_animation.mp4}
\end{figure*}

\begin{figure*}[htbp]
  \centering
  \includegraphics[scale=0.37]{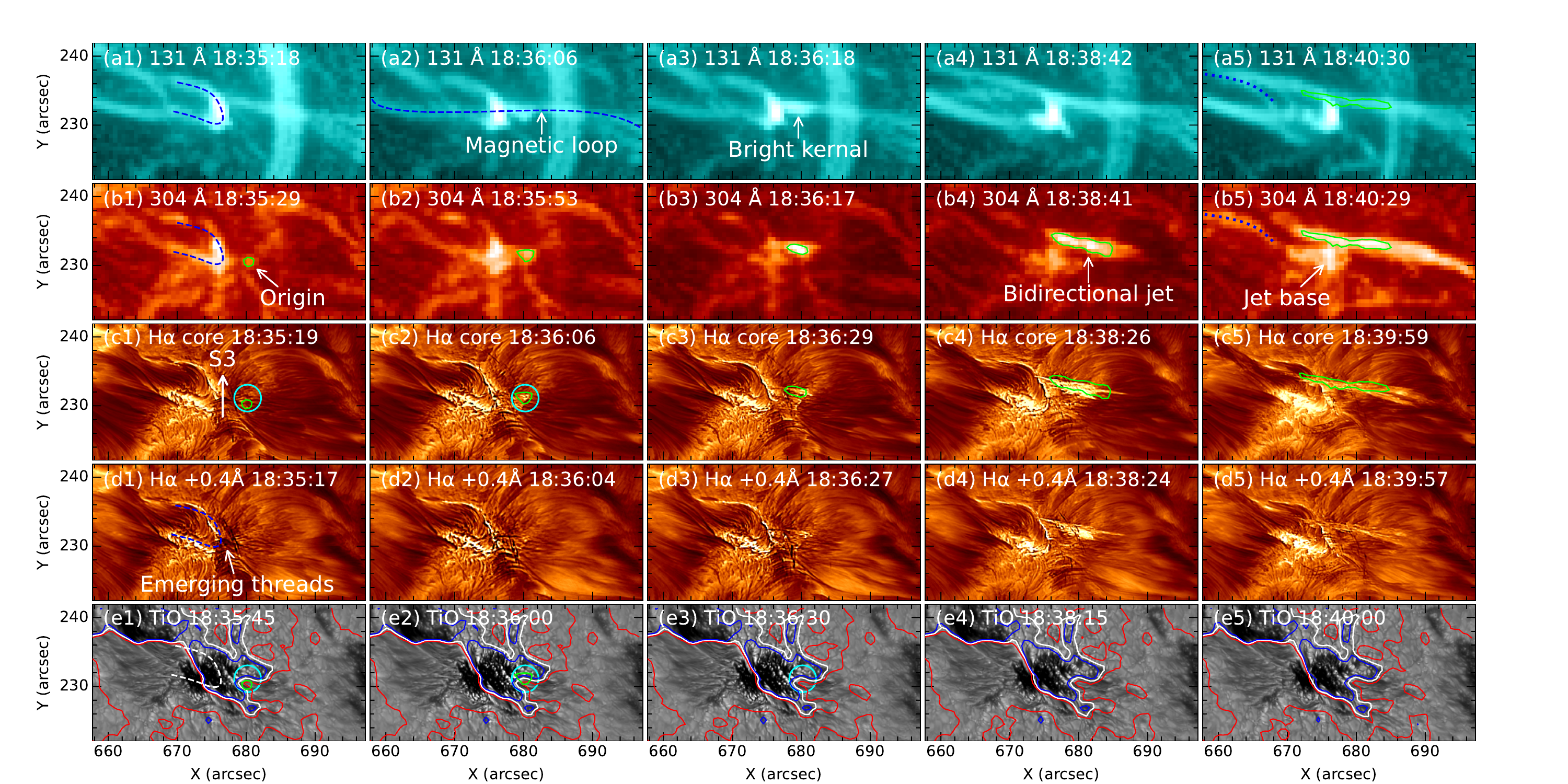}
  \caption{Time-series images in the sub-region observed by AIA and GST. Panels (a-b): The bidirectional jet (JET1) in AIA 131~\AA{} and 304~\AA{} images from 18:35 to 18:40 UT, showing a brightening cusp structure and bidirectional outflow jets. Panels (c-e): A series of chromospheric and photospheric images show the bidirectional jets in H$\alpha$ line core (6562.8~\AA{}), red wing (0.4~\AA{}) and TiO band. The contours with value of $\pm$150 G (red: positive field, blue: negative field) are overlaid on TiO images. The green contours overlaid on background images indicate emission in 304~\AA{} at different time. These circles in panels (c1-c2 and d1-d3) have the same position and correspond to the initial brightening of the jet. The dashed lines S3 and S4 in panels (c1-c2) show the positions marked as spacetime diagrams in Figure 4. An animated version of this figure is available in the online Journal. The animation includes all of the GST images in the static figure (H$\alpha$ line core (6562.8~\AA{}), red wing (0.4~\AA{}) and TiO band) as well as additional H$\alpha$ line wings (-0.4~\AA{}; $\pm$0.6~\AA{}). The animation covers from 18:15 to 19:20 UT, showing two groups jets with high-resolution observations taken by GST. The video duration is 10 seconds.}
  \label{f2}
  \movie[width=0.8\textwidth,showcontrols]{}{f2_animation.mp4}
\end{figure*}

\begin{figure*}[htbp]
  \centering
  \includegraphics[scale=0.38]{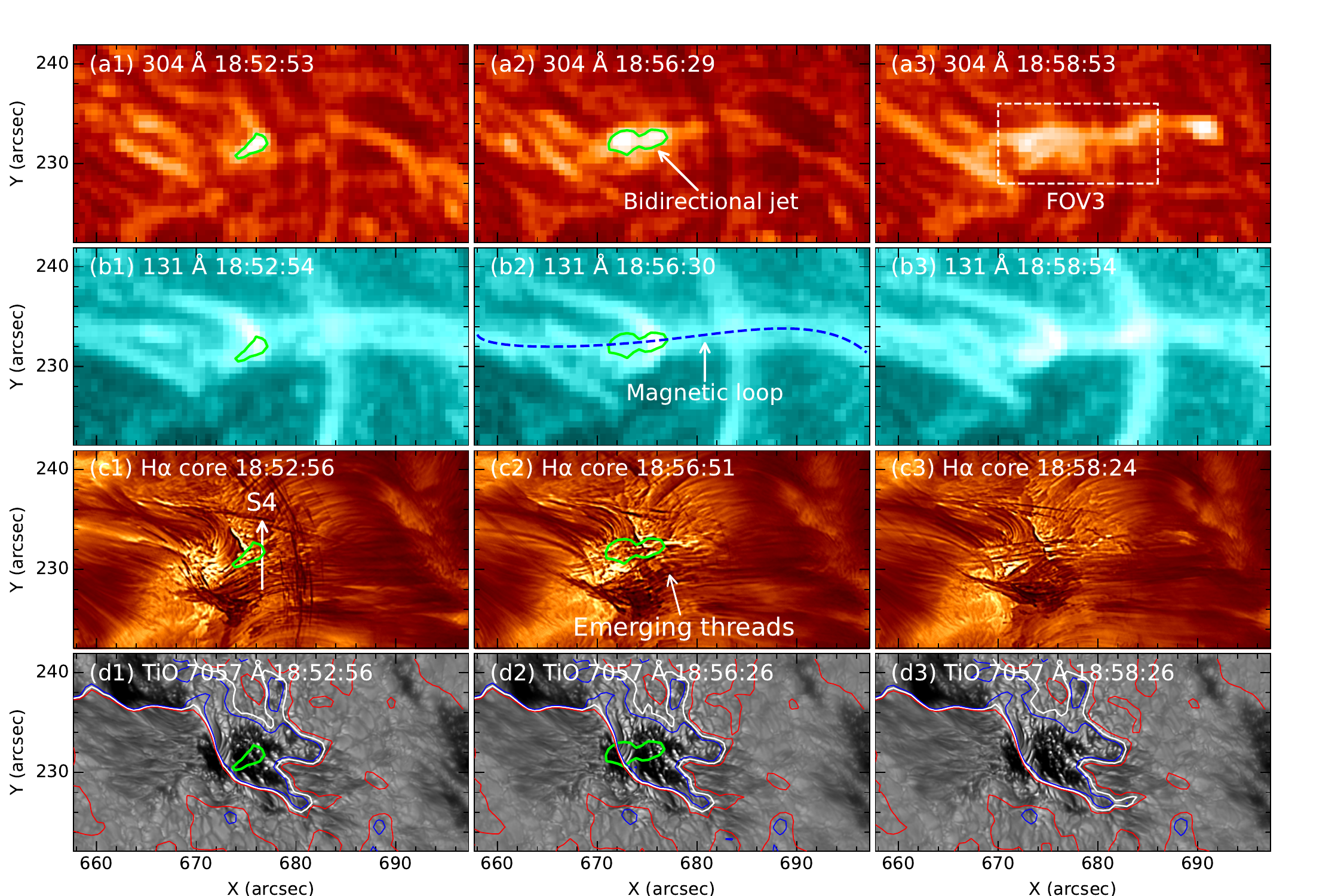}
  \caption{Panels (a-b): The evolution of the second bidirectional jet (JET2) in AIA 304~\AA{} and 131~\AA{} images from 18:54 to 18:59 UT. The white box (FOV3) in panel (a3) indicates the region where the jet's flux was calculated for Figure 6 (a). Panels (c-d): H$\alpha$ line core (6562.8~\AA{}) and TiO band are given from chromospheric and photospheric observations. The contours with the value of 0, $\pm$100 G (white: PILs, red: positive field, blue: negative field) are overlaid on TiO images. The white dashed line (S4) in panel (c1) gives the positions of spacetime diagrams in Figure 4 (c2). The green contours with 90$\%$ maximum emission in 304~\AA{} indicate the initial location of bidirectional outflow jets. }
  \label{f3}
\end{figure*}

\begin{figure*}[htbp]
  \centering
  \includegraphics[scale=0.34]{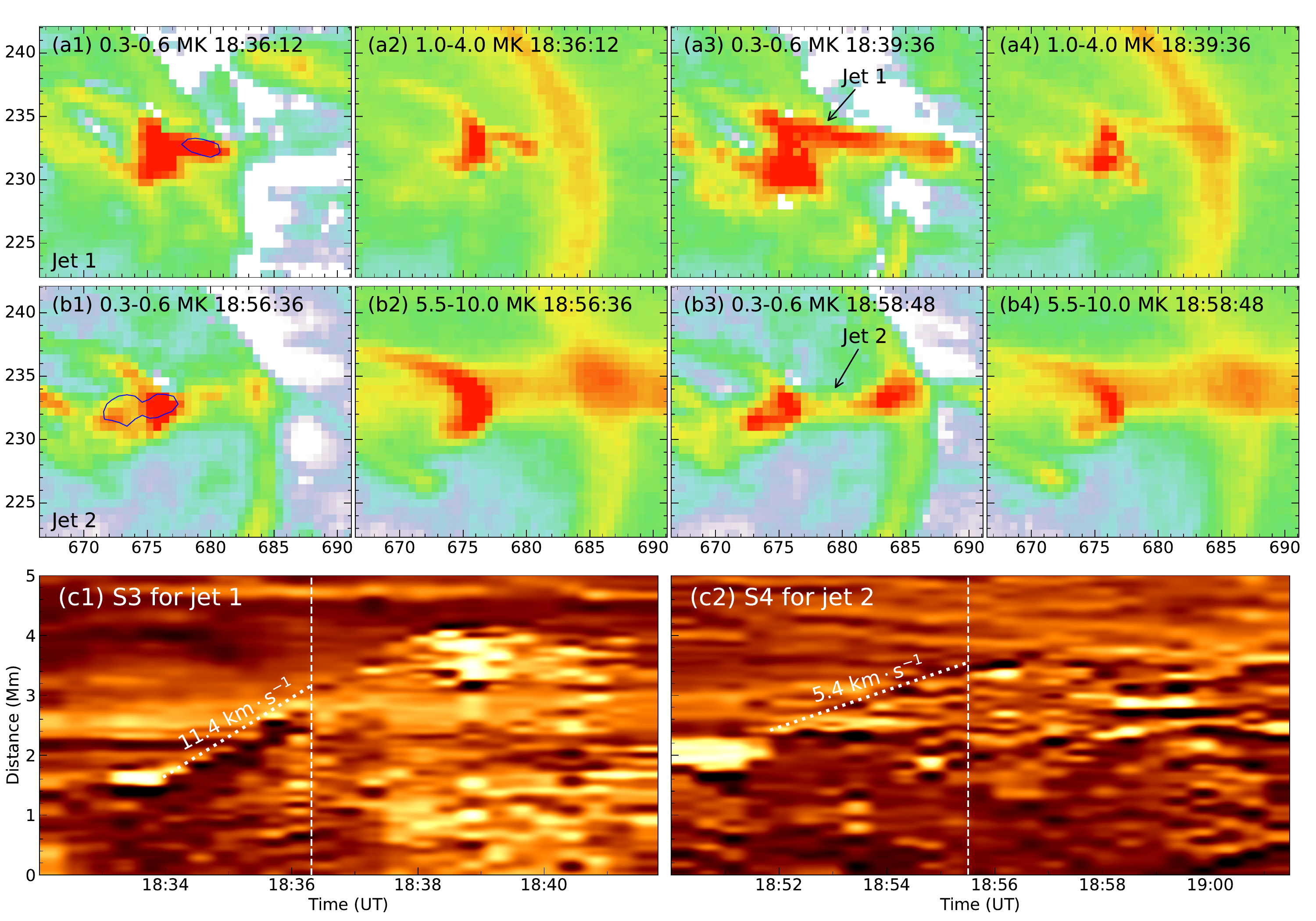}
  \caption{Emission measure and temperature maps obtained by AIA images. Panels (a-b) show the DEM maps of JET1 and JET2 in two temperature ranges. These contours with 90$\%$ of the maximum emission in 304~\AA{} overlaid on images show the initial location of bidirectional outflow jets. Panels (c1-c2) give the space-time images of S3 and S4 in H$\alpha$ line core (6562.8~\AA{}), showing the upward-moving dark threads and the extension of central excitation. The vertical dashed lines correspond to the onset of the two bidirectional jets and show the onset of the magnetic reconnection. The declining dashed lines indicate the moving speeds of the rising filamentary threads.}
  \label{f4}
  
\end{figure*}

\begin{figure*}[htbp]
  \centering
  \includegraphics[scale=0.4]{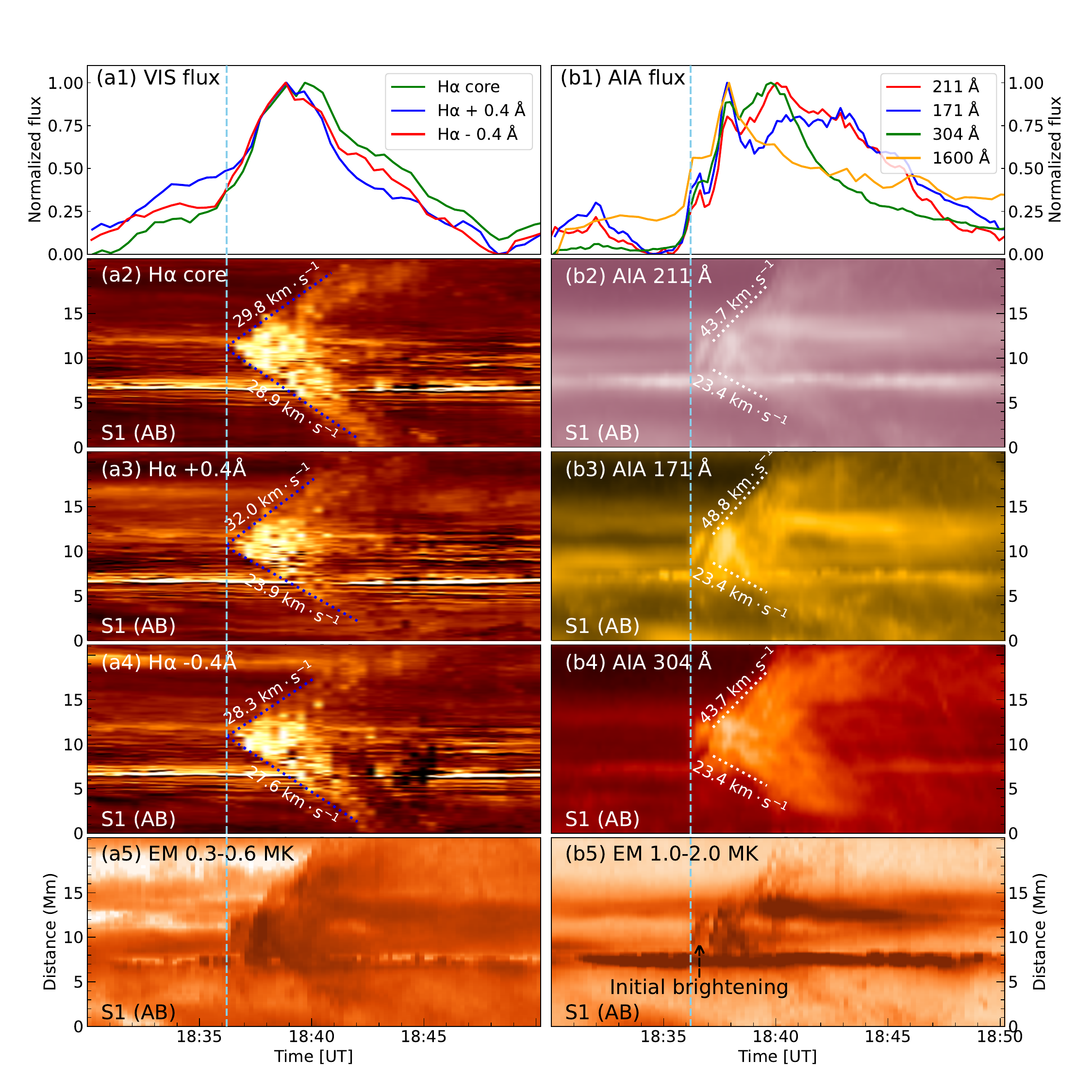}
  \caption{Panels (a1-b1) show the normalized light curves in H$\alpha$ and EUV wavelengths during the first jet. Panels (a2-a4) and (b2-b4) display the space-time diagrams along S1 (JET1) in H$\alpha$ line core (6562.8~\AA{}), wing ($\pm$0.4~\AA{}, AIA 211, 171, and 304~\AA{}, showing a bidirectional outflow jet. The temperature evolution of the jets shows the similar results with the EUV images in panels (d1-d2). The blue dashed lines indicate the onset of the bidirectional plasma jet. The declining dashed lines indicate the moving speeds of the bidirectional jets.}
  \label{f5}
\end{figure*}

\begin{figure*}[htbp]
  \centering
  \includegraphics[scale=0.35]{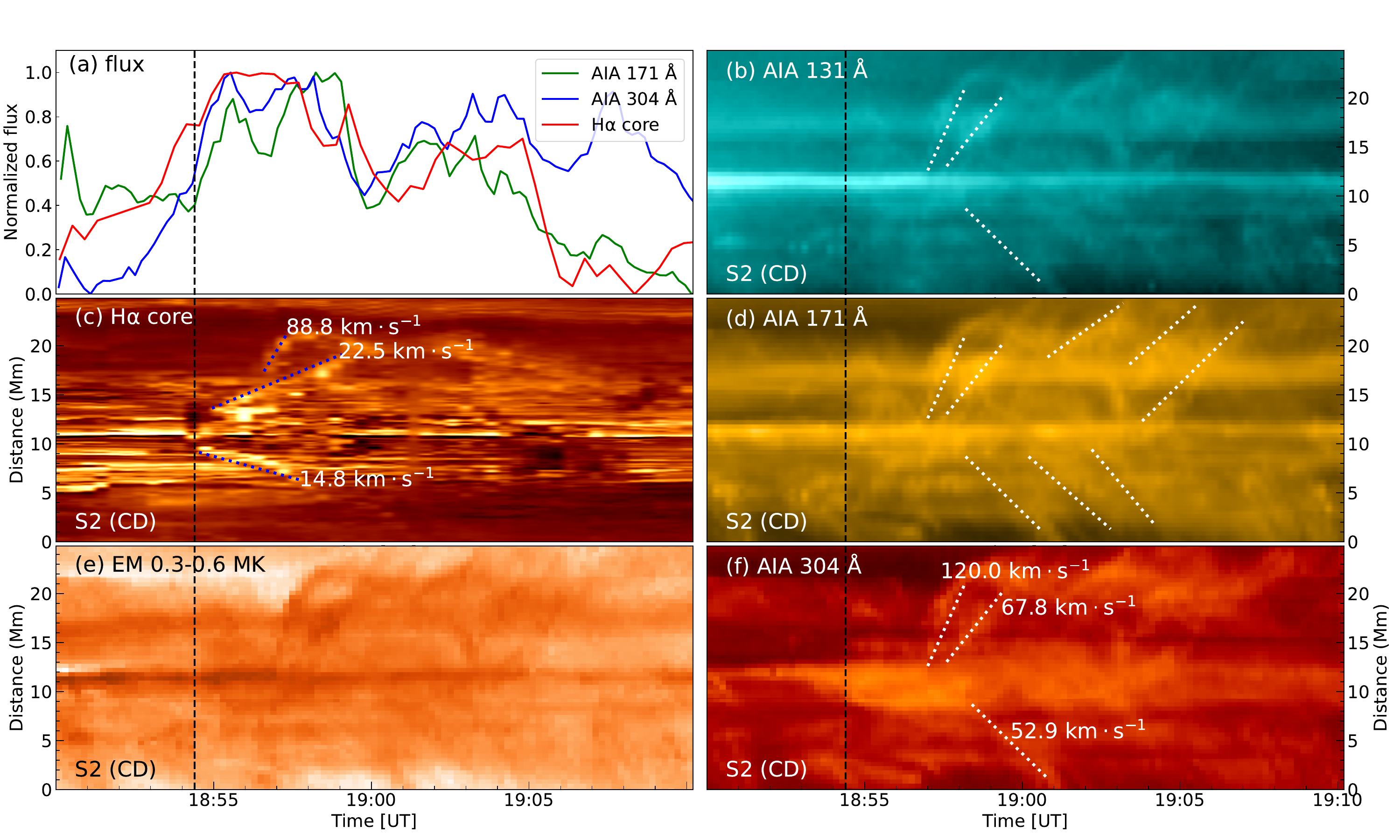}
  \caption{Spacetime diagrams and the light curves during the JET2. Panel (a) shows the normalized light curves during the JET2 in H$\alpha$ line core and AIA 304 ~\AA{} and 131 ~\AA{}. Panels (b-d) and (f) display spacetime diagrams along S2 in H$\alpha$ line core, AIA 131, 171 and 304 ~\AA{} respectively. The temperature evolution of the bidirectional jet is shown in bottom panel  (e). The black dashed lines indicate the onset of the second bidirectional jet. The declining lines are the moving speeds of the jets.}
  \label{f6}
\end{figure*}

\begin{figure*}
  \centering
  \includegraphics[scale=0.9]{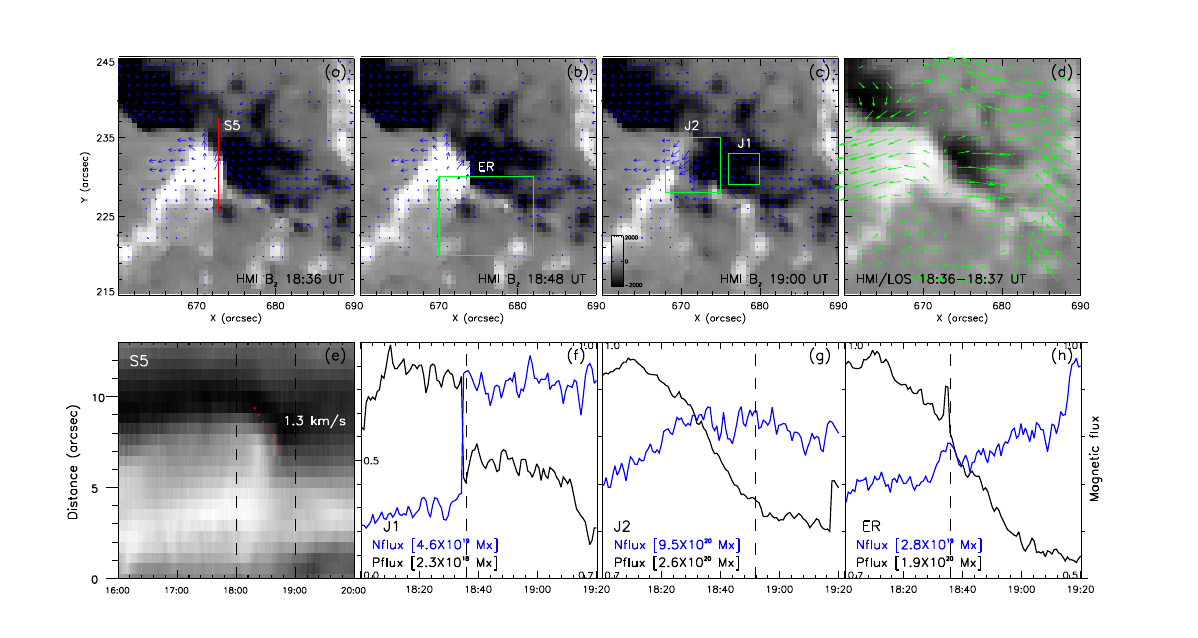}
  \caption{Panels (a-c) show the evolution of magnetic field in the localized active region, the blue arrows are the horizontal magnetic field with a strength greater than 300 Gauss. Panels (d-e) show the velocity field and the spacetime diagram along S5, and the green arrows represent the LCT flow map in the small region obtained during 18:36–18:37 UT. The three green boxes are the region that we used to calculate the magnetic flux, corresponding to the photospheric regions of JET1, JET2 and emerging region. Panels (f-h) display the evolution of positive and negative magnetic flux in the three regions. The dashed lines indicate the onset of the two groups of bidirectional jets.}
  \label{f7}
\end{figure*}

\section{Results}
To study the dynamic evolution of the two jets, we applied the multi-Gaussian normalization (MGN) method \citep{morgan2014multi} to obtain enhanced images in AIA 94, 131, 304, and 335 ~\AA{} wavelengths. For our analysis, only the AIA imaging observations were used for the MGN enhancement, while DEM analysis was applied on EUV data without MGN enhancement. In Figure 1, the two successive bidirectional plasma jets (named after JET1 and JET2) are observed in the active region. For the JET1, its extension length reaches $40^{\prime\prime}$, lasting from 18:35 to 18:43 UT, and JET2 is approximately $30^{\prime\prime}$ and ranges from 18:55 to 19:05 UT. Figure 1 (a-d) give the first group of bidirectional jets in different wavelengths at $\sim$ 18:40 UT, a jet base and bidirectional jets can be observed in the lower temperature 304~\AA{} (90$\%$ of the maximum emission in panel (c)), with newly formed magnetic loops. According to these panels and their animation, the cusp structure (blue curve lines in panels (b) and (f)) originates from the light bridge-like penumbra located in the vicinity of PILs, and shows a change from penumbra to granular during the two jets. At the same time, continuous erupting plasma blobs (in brightening cusp) are visible in cool and hot channels from 18:25 to 18:32 UT, and lead to the heating of surrounding plasma and magnetic loops (white arrows in panel (b)). Combined with the animation of this figure, a large amount of the filament material and filamentary threads continuously emerge, the two successive bidirectional plasma jets are generated through the bipolar emerging flux and the overlying horizontal fields. 

Figure 2 shows a sequence of temporal and spatial evolution of the JET1 in the sub-region. The first two rows display its evolution in AIA 131~\AA{} and 304~\AA{} at different time. At $\sim$ 18:35 UT, a bright kernel first appears in 304~\AA{}, then can be observed in magnetic loops of 131~\AA{}, after that it generally evolves into the bidirectional jets due to ongoing magnetic reconnection. Simultaneously, the filament materials are slowly rising and restrained inside the heating magnetic loops, and visible in the animation of Figure 1. According to the EUV observations and its animation, the bidirectional plasma jets not only extend from the central excitation location in opposite directions but also show an upward motion, due to continuous magnetic reconnection between the rising filament and overlying horizontal magnetic loops.

Panels (c-d) show H$\alpha$ images in line core and red wing (+0.4~\AA{}) from high-resolution GST observations. It can be seen that many parallel dark threads are moving upwards and continuous emerging. According to the report by \cite{bjorgen2019}, these filamentary threads show absorption features produced by the radiative transfer process of the first line of the Balmer series, which are believed to be chromospheric fibrils and mostly aligned with the magnetic field direction. At $\sim$18:36 UT, the central brightened kernel is observed while the filament-like strands are interacting with the horizontal magnetic field above it, which is believed to be the location where magnetic reconnection occurs. It is noted that bidirectional jets show a continuous upward movement and gradually turn into the splitting filamentary threads. Compared with AIA 304~\AA{} images, the initial brightening in H$\alpha$ wavelength is delayed for approximately 30 seconds, indicating that the reconnection happens at a higher altitude (the transition region). During the JET1, a large number of nanoflare-like intensity bursts are also observed in H$\alpha$ wavelengths at 18:40 UT. Simultaneously, the heated magnetic loops show a braided to almost parallel morphology within a few minutes, followed by the outflow plasmoids along the loops, visible in AIA 304 and 131~\AA{}. For the first column and its animation, the brightening cusp structure marked as curves corresponds to the brighten ribbon of H$\alpha$. Simultaneously, the initial brighten (see the cyan circle) appears in the intersection of umbra and penumbra of the sunspot located at PILs, indicating the magnetic emergence and cancellation in the lower atmosphere.

Figure 3 shows the second group of bidirectional jet (JET2) at different wavelength (from top to bottom in sequence: 131~\AA{}, 304~\AA{}, H$\alpha$ core and TiO band images). According to EUV images and animation of Figure 1, these filament materials are initially confined inside the magnetic loops. At $\sim$18:53 UT, a brightening point (the initially jet) suddenly appears in  the region due to the magnetic cancellation. When the filament materials rise rapidly and interact with the heating magnetic loops, this leads to the central excitation and heating of the surrounding plasma. The bidirectional jets in opposite directions are formed due to the magnetic reconnection between them. During the animation for JET2, many of small intensity bursts first occur inside the rising filamentary threads, and erupting plasmoids along the heated magnetic loops. It is found that the heated coronal magnetic lines transform from
a crossing state into an almost parallel morphology between 18:58 and 19:07 UT. Compared with EUV observations, the formation of the bidirectional jets of H$\alpha$ is due to the interaction between the emerging filamentary threads with twisted structure and overlying magnetic fields. For the JET1 and JET2, the background magnetic fields where magnetic reconnection occurs are located in the different height. Through photospheric observations in TiO band and magnetographs, the initial brighten of JET2 is located in the bright bridge like penumbra along PILs, it can be seen that the photospheric granular gradually form and begin to expand during the jet, which are simultaneously accompanied by the magnetic flux emergence and cancelations in the localized region. In addition, it is found that these filamentary threads are along PILs and are emerging from the lower atmosphere.  


In order to obtain the temperature and density-dependent emission of the two bidirectional jets, we carried out differential emission measure (DEM) analyses and obtained the distributions of the EM and temperature using the original data from AIA six wavelengths. The top row of Figure 4 demonstrates the selected temperature maps of JET1 during its eruption, temperature range for the integrated EM are from 0.3 MK to 4 MK, while the JET2 in the middle row shows the higher emission and temperature (until 10 MK) than the JET1, this might be due to the fact that the second group of jets are not only recurrent but also confined within magnetic loops, no showing the rapid ascending. In addition, the jet base of JET1 and JET2 has the stronger emission than its spires.

We made the spacetime diagrams of JET1 and JET2 (along S3 and S4) that are perpendicular to the rising filamentary threads. According to the space-time diagrams of Figure 4, it can be obtained how and where magnetic reconnection occurs. From animation of Figure 2 and panels (c1-c2), it further confirmed that the successive bidirectional jets are caused by the interaction between the continuously ascending filamentary threads and the overlying horizontal magnetic field. For the JET2, the height and position of magnetic reconnection are different from the JET1 because the filaments in the second group have a highly twisted structure. The rising speeds of filamentary threads are about 11.4 and 5.4 ~\text{km~s$^{-1}$} for JET1 and JET2 respectively, and the magnetic reconnection between them occurs when the dark threads stop rising and interact with overlying magnetic fields. For the JET1 and JET2, the location of reconnection rises a height of about $3-4^{\prime\prime}$ and lasts about 10 minutes. In addition, the reconnection between the ascending filamentary threads generates many discrete bursts, similar to the reconnection nanojets reported by \cite{antolin2021}.

Figure 5 displays the light curves (the small rectangular box of Figure 1 (d)) and spacetime diagrams of the JET1. Based on high-resolution H$\alpha$ observations of GST, we obtain the spacetime diagrams along S1 path in H$\alpha$ line core (6562.8~\AA{}) and wing ($\pm$0.4~\AA{}), showing a symmetrical bidirectional jet from the central excitation brightening, looks like butterfly images, and EUV's observations also display similar behaviour. Meanwhile the initial EUV brightening kernels in spacetime diagrams are also observed and prior to H$\alpha$, there is a delay time of 30 s between them, similar to the images in Figure 2. From these spacetime diagrams, the brightening ribbons (corresponding to the cusp structure) are located below the position of magnetic reconnection (about $3^{\prime\prime}$). And the bidirectional jet in H$\alpha$ and EUV observations lasts for approximately 10 minutes. Since H$\alpha$ images taken by GST provide the sub-arcsec spatial resolution, it can be seen that there are three periodic erupting plasma jets from magnetic reconnection, but are invisible in EUV observations. Furthermore, filamentary material is not only continuously emerging and rising before the onset of the jets but also confined inside the magnetic loops during the two jets. 

Spacetime diagrams of JET2 (along S2) are shown in Figure 6. It is found that an initial jet is next to the cusp structure (see H$\alpha$ line core and AIA 304~\AA{}), and a symmetrical bidirectional plasma jet is only observed in H$\alpha$ line core. Compared with H$\alpha$ observations, the JET2 first shows approximately 3 minutes of brightening and extension in 304~\AA{}, then showing repeated recurrent in different EUV bands. It is very likely that the bidirectional jet from H$\alpha$ and EUV observations should occur in different altitude. At $\sim$ 18:56 UT, another jet is observed in AIA and H$\alpha$ simultaneously. During the two bidirectional jets, there are a few obvious observing features: 1) initially brightening occurs in transition or chromosphere region and locates at PILs; 2) the bidirectional jets from reconnection site show the movement in opposite direction; 3) the magnetic reconnection occurs between the rising filament threads and the overlying magnetic loops and shows an upward motion; 4) a large amount of micro eruptions are generated due to the reconnection between filamentary threads, and are accompanied by the heated coronal magnetic loops. These observations are consistent with the magnetic flux emerging model described in the 2D numerical model \citep{yokoyama1995} and observations of bidirectional jets driven by eruptive filament \citep{sterling2019}. 

The normalized integrated intensity from different wavelengths show obvious enhancement during the two bidirectional jets in Figure 5 and Figure 6. According to the spacetime diagrams, we calculate the average speeds of the bidirectional eruptions to be 21.6 and 48.8 ~\text{km~s$^{-1}$} for JET1, respectively. The average ejection speeds of JET2 range from 14.8 to 120 ~\text{km~s$^{-1}$}. In addition, Figure 5 (a5-b5) and Figure 6 (e) show the spacetime diagrams of temperature along S1 and S2. It can be seen that the DEM also displays a similar bidirectional distribution during the JET1 and JET2.
 
The evolution of vertical magnetic fields ($B_{z}$) during the two jets is shown in Figure 7. For the two localized regions with bidirectional jets (JET1 and JET2), the magnetic fields show a obvious cancellation along PILs, it is can be seen that the negative magnetic fields generally invade into the positive region. The positive magnetic fields are decreasing, while the negative fields are increasing with time. And horizontal magnetic fileds $B_{h}$ (blue arrows) not only enhance but also show an obvious change in the core region, their moving direction first spreads outward, then toward positive fields. Figure 7 (d) displays the flow field vectors derived from the small region over 18:36–18:37 UT using the local correlation tracker (LCT) method \citep{fisher2008flct}. It shows the shearing and converging flows along PILs in the {$\delta$ }sunspot region with jets. Panel (e) gives the spacetime diagram along S5, the negative magnetic field enters the positive field with a speed of 1.3 ~\text{km~s$^{-1}$}.In addition, panels (f-h) show the temporal profiles of the integrated magnetic fluxes in the two jets and emerging region of filamentary threads. During the JET1 and JET2, we find that the positive and negative magnetic fluxes are rapidly decreasing and increasing respectively, indicating obvious magnetic cancellation. For the emerging region, there is a general ﬂux decrease over the two jets, however, a prominent jump of about 5 minutes begins nearly exactly with the onset of the eruption of filamentary threads and the JET1.

\section{Conclusion and discussion}

In this paper, we present the formation of the bidirectional plasma jets driven by magnetic reconnection from observations of AIA and GST. The two successive bidirectional jets are caused by the interaction between the rising filamentary threads or filament material and overlying horizontal magnetic loops. The initial brightening located at PILs occurs first in the transition region or chromosphere, and followed by the emergence and cancellation of magnetic flux. Our main results are as follows.

According to the evolution of the magnetic fields, we find that the positive and negative magnetic fields show simultaneous decreases and increases during JET1 and JET2. Meanwhile, the negative magnetic field enters the positive field with a speed of 1.3 ~\text{km~s$^{-1}$}, showing obvious magnetic cancellation in PILs. The horizontal fields not only show significant enhancement, their movement directions have also changed along PILs. For photospheric observations in TiO band and magnetographs, the two jets initially appear in the interface of the penumbra and umbra of a sunspot and are located in the vicinity of PILs. These results indicate that magnetic emergence and cancellation play a major role in powering the continuous bidirectional plasma jets.

From the chromospheric and coronal observations (H$\alpha$ and EUV) during JET1 and JET2, we find that these filamentary threads with horizontal direction continuously emerge before JET1 and JET2, and this is considered to be the initial filamentary material. As the filament material begins to rise and reconnect with the horizontal magnetic loops, the ongoing magnetic reconnection appears between them. Actually, reconnection between the rising filament (more speciﬁcally, holding the cool material) and those horizontal magnetic loops results in localized heating at the reconnection location and forms the bidirectional jets, and accompanied by the erupting plasmoids along the heated magnetic loops.

The ultra-fine filamentary threads are emerging from the sunspot penumbra. We find that the initially filamentary threads are generally transform from a almost-parallel into a twisted a twisted configuration during the JET2, which may be due to the evolution of photospheric field and photospheric granular. In addition, according to the animation of Figure 1, high-resolution H$\alpha$ observations show a large number of nanoflare-like intensity bursts during the two jets. And we find that the heated coronal magnetic lines transformed from a twisted state into an almost parallel morphology, accompanied by the erupting plasmoids along magnetic loops.

Unlike JET1, the JET2 shows repeatedly recurrent jets in opposite directions, which may depend on the structure of rising filament threads and background magnetic fields. And the average speeds of the bidirectional eruptions range from 14.8 to 120.0 ~\text{km~s$^{-1}$}. In addition, the initial brightening of the JET1 is first observed in 304~\AA{} ($\lg T \approx 4.7$), prior to the chromospheric observations in H$\alpha$, indicating magnetic reconnection occurs at the height of transition region. We thought that generation of bidirectional plasma jets is due to the magnetic reconnection between the rising filamentary threads (or filament material) and the overlying magnetic loops.

Recent observations have reported that two-sided loop jets are caused by the eruption of a miniature filament \citep{sterling2019,shen_Stereoscopic_2019,yang2024}. Figure 7 of \cite{sterling2019} shows the physical picture of bidirectional jets powered by the external magnetic reconnection, which originates from the rising minifilament field reconnecting with the overlying horizontal field, while inside reconnection between the legs of twisted filament lead to the brightening loop of jet base, consistent with the observational results reported by \cite{yang_B_2019}, nevertheless, in their paper, the bidirectional jets are caused by magnetic reconnection between erupting minifilaments and the overlying large-scale filament. \cite{tian2017} thought that the two successive bidirectional jets result from the interaction of two approximately parallel and adjacent filamentary threads, rather than the magnetic flux emergence model as reported by \cite{shibata1994a} and \cite{jiang2013}, which is driven by interaction between the emerging bipolar and the overlying horizontal magnetic fields. For our case, the two bidirectional jets are due to the magnetic reconnection between the rising filament threads and newly formed magnetic loops. In addition, recurrent two-sided loop jets are also reported by some observations \citep{jiang2013, tian2017, yang2019, yan2021}, that is due to the continuous interacting emergence of magnetic flux with the overlying horizontal magnetic field.

Using three dimensional magnetohydrodynamical (MHD) simulations, \cite{wyper2017, wyper2018} reported that the jet driven by mini-filament eruption is due to the magnetic reconnection between the emerging magnetic flux and the external opening field. In this case, a filament channel strongly sheared is formed at PILs due to the intrusion between the positive and negative polarity concentration. Here, during the JET1 and JET2, we find that the strong negative field invaded the positive field, leading to the strongly sheared at the polarity inversion line and enhancement of the horizontal field, similar to the simulation reported by \cite{wyper2017}. In addition, hot and cool components for the two bidirectional jets could not be observed to coexist. It can be found that the bidirectional jets in H$\alpha$ observations gradually become the longer and slimmer filamentary threads, which are different from the reported work of \cite{shen2021}. In summary, using the high-resolution observations from the GST and AIA, we find that the initial filamentary threads and filament material are activated and gradually form the large-scale filament, which are closely related to the bidirectional jets and subsequent flaring eruptions, respectively. Our observations show the formation process of bidirectional jets driven by magnetic reconnection from lower to higher atmosphere. 

According to high-resolution observations and numerical simulations, \cite{antolin2021} reported that a large number of localized reconnection between misaligned coronal magnetic field lines at small angles leads to the bidirectional nanojets. During the expanding coronal loops, they found that the cluster nanojets not only show a movement that is perpendicular or parallel to the reconnecting field lines but also result in heating of the coronal magnetic loops. Therefore, the nanojets play an important role in reconnection-based coronal heating. In our work, the large-scale bidirectional jet shows a horizontal plasma outflow in opposite direction and is accompanied by the heating magnetic loops. Meanwhile, during the two jets, a large number of nanoflare-like intensity bursts are observed in H$\alpha$ core line, visible in the time-space diagrams and the animation of Figure 2. And the heating magnetic loops almost simultaneously occur and show a crossing state to almost-parallel morphology, with a lasting nanoflare (the cusp structure), which might further lead to plasma heating and bidirectional outflowing plasmoids (see animation of Figure 1), consistent with the report by \cite{antolin2021} and \cite{chen2025}. However, it cannot provide the evolution of these small-scale explosions due to the limitations of resolution and the observational background. There is no doubt that high-resolution observations are extremely important for understanding the drivers of nanojets and reconnection-based coronal heating.

\begin{acknowledgements}{We would like to thank the anonymous reviewer for the constructive comments that have enabled us to improve the manuscript. SDO data were made available by the NASA/SDO AIA and HMI science teams. We are grateful to the teams of SDO and GST for providing high-quality data. This work is supported by National Natural Science Foundation of China (NSFC) under grants 12273101. The work is supported by the Strategic Priority Research Program of the Chinese Academy of Sciences Grant No. XDB0560000. The work is also sponsored by Natural Science Foundation of Xinjiang Uygur Autonomous Region under grant 2024D01E38, Funded by the Xinjiang Talent Development Fund XJRC-2025-KJ-YJ-CXPT-066XJRC-2025-KJ-YJ-CXPT-066, and XJRC-2025-KJ-YJ-CXPT-179. BBSO operation is supported by US NSF AGS-2309939 grant and New Jersey Institute of Technology.}

\end{acknowledgements}

\bibliography{bipolar_ref}{}
\bibliographystyle{aasjournalv7}

\end{document}